\documentclass{article}

\usepackage{arxiv}

\usepackage[utf8]{inputenc} 
\usepackage[T1]{fontenc}    
\usepackage{hyperref}       
\usepackage{url}            
\usepackage{booktabs}       
\usepackage{amsfonts}       
\usepackage{nicefrac}       
\usepackage{microtype}      
\usepackage{lipsum}		
\usepackage{graphicx}
\usepackage[numbers]{natbib}
\usepackage{doi}

\usepackage{lineno}
\usepackage{mhchem}
\usepackage[utf8]{inputenc}
\usepackage{tabularx}
\usepackage[table]{xcolor} 

\title{Polarization-Multiplexed Bloch Surface Wave Sensing of Single-Strand DNA Growth}

\author{
  Jonathan Barolak\thanks{These authors contributed equally to this work} \\
  Dipartimento di Fisica\\
  Politecnico di Milano\\
  20133 Milano, Italy \\
  \And
  Erika Mogni\footnotemark[1] \\
  Dipartimento di Fisica\\
  Politecnico di Milano\\
  20133 Milano, Italy \\
  \And
  Giovanni Pellegrini \\
  Dipartimento di Fisica\\
  Universit\`a degli Studi di Pavia\\
  27100 Pavia, Italy \\
  \And
  Jorge Gil-Rostra \\
  Instituto de Ciencia de Materiales de Sevilla\\
  CSIC-Universidad de Sevilla\\
  41092 Sevilla, Spain \\
  \And
  Francisco Yubero \\
  Instituto de Ciencia de Materiales de Sevilla\\
  CSIC-Universidad de Sevilla\\
  41092 Sevilla, Spain \\
  \And
  Michele Celebrano \\
  Dipartimento di Fisica\\
  Politecnico di Milano\\
  20133 Milano, Italy \\
  \And
  Marco Finazzi \\
  Dipartimento di Fisica\\
  Politecnico di Milano\\
  20133 Milano, Italy \\
  \And
  Katharina Schmidt \\
  Laboratory of Life Science Technologies\\
  Danube Private University\\
  Wiener Neustadt, Austria \\
  \And
  Stefan Fossati \\
  FZU-Institute of Physics\\
  Czech Academy of Sciences\\
  Prague, Czech Republic \\
  \And
  Jakub Dost\'alek \\
  Laboratory of Life Science Technologies\\
  Danube Private University\\
  Wiener Neustadt, Austria \\
  \And
  Paolo Biagioni \\
  Dipartimento di Fisica\\
  Politecnico di Milano\\
  20133 Milano, Italy \\
  \texttt{paolo.biagioni@polimi.it} \\
}

\renewcommand{\shorttitle}{\textit{arXiv} Template}

\begin{document}
\maketitle

\begin{abstract}
Refractometric biosensing is a vital label-free tool for the real time detection and interaction analysis of biological and chemical substances. Nanophotonic platforms like Surface Plasmon Resonance (SPR) have played a critical role in providing refractometric sensing capabilities for clinical diagnostics and environmental monitoring. However, traditional systems operating in a single-polarization state cannot fully characterize complex optical properties such as birefringence, which is crucial to resolve many complex biological interactions. Although Bloch Surface Wave (BSW) sensors can support both Transverse Electric (TE) and Transverse Magnetic (TM) modes, a key capability SPR lacks, they have historically been implemented in single-mode configurations. In this paper, we present a polarization multiplexed BSW refractometric sensing system, simultaneously tracking the resonant wavelength shifts of both TE and TM BSW modes through time. Our technique was applied to investigate single-strand DNA growth during rolling circle amplification (RCA). To accurately recover the time-dependent birefringence, capturing dynamics of the DNA growth and orientation of its chains, we implemented a two-stage modeling approach based on the Transfer Matrix Method (TMM). First, we utilized a wavelength-dependent surface sensitivity model, confining refractive index changes to the immediate layer above the crystal, to distinguish isotropic background dynamics from birefringent signals. Following the onset of RCA, we transitioned to a model that accounted for the vertical growth of the DNA layers in time. By fitting this model to the TE and TM resonant shifts, we monitor the growth rate of the single-strand DNA layer as well as the refractive index along the two polarization components. Our findings demonstrate the platform’s ability to resolve the structural evolution of complex bimolecular interactions associated with conformational changes.
\end{abstract}

\section*{Introduction}

Accurate and rapid determination of a material’s refractive index is essential in technologies ranging from chemical and gas detection to environmental monitoring and point-of-care diagnostics \cite{grygaBlochSurfaceWave2020,capelliSurfacePlasmonResonance2023,descroviCouplingSurfaceWaves2007}. Nanophotonic platforms capable of such measurements have matured into a powerful class of refractometric sensors, offering high sensitivity and low detection limits through the combination of strong optical confinement and biochemical surface functionalization for analyte pre-concentration \cite{arrudaUltrasensitiveAlzheimersDisease2025}. Among these, Surface Plasmon Resonance (SPR) sensors have become widely used due to their simple fabrication and optical configurations \cite{homolaSurfacePlasmonResonance2008,svedendahlRefractometricSensingUsing2009,prabowoSurfacePlasmonResonance2018}. By tracking the resonant optical signal associated with a surface plasmon (SP), refractive index dynamics can be monitored in real time without physical contact \cite{kabashinPhaseAmplitudeSensitivities2009,svedendahlRefractometricBiosensingBased2014}. SPR based biosensors have played a transformative role in life sciences and clinical diagnostics, enabling rapid, quantitative, label-free analysis of biomolecular interactions \cite{steglichSurfacePlasmonResonance2022}. However, SPR sensors are restricted to a specific linear polarization, limiting their capacity to probe birefringent or anisotropic dynamic systems.

Over the last decade, dielectric nanophotonic platforms such as Bloch Surface Waves (BSWs) have emerged as a compelling alternative to SPR sensing platforms \cite{michelottiBlochSurfaceWaves2025}. BSWs, the dielectric analog of SPs, are generated with One Dimensional Photonic Crystals (1DPCs), which confine light at the surface through the combined effects of total internal reflection and the presence of a photonic bandgap \cite{occhiconeSpectralCharacterizationMidInfrared2021a,aurelioElectromagneticFieldEnhancement2017a,michelottiBlochSurfaceWaves2025}. BSWs can be designed across a broad spectral range, exhibit significantly lower absorption losses compared to SPs, and, crucially, support both Transverse Electric (TE) and Transverse Magnetic (TM) polarizations states \cite{khanBlochSurfaceWave2016}.  Additionally, reports have shown that optimized BSW sensors can actually exhibit stronger field confinement and improved sensing characteristics compared to SPR based sensors \cite{lereuSurfacePlasmonsBloch2017,rizzoBlochSurfaceWave2018,sinibaldiDirectComparisonPerformance2012}. These properties have made BSWs increasingly attractive as an alternative to SPR for refractometric sensing. Though many studies have investigated and applied the enhanced characteristics of BSW sensing to various biosensing scenarios \cite{sinibaldiBlochSurfaceWaves2017,rizzoBlochSurfaceWave2018}, they have focused on sensing with only a single polarization state.

Simultaneous excitation/monitoring of TE and TM BSWs could enable polarization-resolved measurements of surface refractive index dynamics. Such a system would allow for direct access to the time-dependent birefringence of thin films and interfacial layers, an effect that is particularly relevant in anisotropic biomolecular systems such as deoxyribonucleic acid (DNA), collagen, or lipid membranes \cite{samocRefractiveindexAnisotropyOptical2007,mishimaOpticalBirefringenceMultilamellar1996,keikhosraviRealtimePolarizationMicroscopy2021}. In practice, however, achieving this condition presents several challenges. The inherent anisotropy of 1DPCs leads to distinct dispersion relations for TE and TM modes, meaning simultaneous excitation of TE/TM modes at the same wavelength and incident angle $(\lambda,\theta_{inc})$ is not generally possible. This precludes the ability to study birefringent dynamics in the common sensing configuration wherein a single photodiode detects the change in amplitude or phase of the reflected beam at a particular $(\lambda,\theta_{inc})$ coordinate \cite{konopskyPhotonicCrystalSurface2007a}. When measurements are performed spectrally, by either recording a wavelength spectrum or an angular spectrum and measuring the shift in resonant wavelength/angle, the perfect overlap of TE/TM modes is not strictly required. However, large discrepancies between the dispersion relations of the TE/TM modes can require a significant wavelength/angular bandwidth of the detection system. By engineering the respective dispersion relations to be spectrally aligned, the walk-off between TE and TM resonances can be minimized, ensuring stable dual-polarization operation over a broad spectral and refractive index range. Designing such 1DPCs is challenging, however superchiral BSW systems have been investigated to design exactly such a system \cite{pellegriniChiralSurfaceWaves2017}. By adding a properly designed layer to the surface of a 1DPC, the TE/TM dispersion relations can be aligned, allowing for broadband, simultaneous excitation of dual polarization BSWs \cite{mogniOneDimensionalPhotonicCrystal2022}. Recent work has demonstrated strategies to reduce the intrinsic anisotropy of 1DPCs and improve TE/TM alignment using genetically optimized low refractive index contrast 1DPCs, thereby enhancing the signal-to-noise ratio balance and sensitivity uniformity between the two polarization channels \cite{barolakLeveragingLowIndex2025}, although the use of low index contrast materials generally lowers the refractometric sensitivity of the BSW device.

In this work, we present a polarization multiplexed refractometric sensor capable of simultaneous excitation/detection of TE and TM polarized BSW modes. We apply this approach to monitor the anisotropic optical response of single-strand DNA (ssDNA) grown via rolling circle amplification (RCA), a representative example of a dynamic, birefringent biomolecular process. We present a method for modeling the wavelength-dependent surface sensitivity of the TE/TM BSW modes. With this numerically determined sensitivity, we convert the experimentally measured resonant wavelength shifts as a function of time to the temporally resolved birefringence of the thin layer just above the surface of the 1DPC. In doing so, we distinguish between isotropic refractive index variations and birefringent dynamics. During the RCA process, we transition from a sensitivity model which assumes refractive index variations within a thin surface layer to a model that accounts for the vertical growth of a birefringent layer. With this model, we fit the experimental data using a nonlinear least squares optimization procedure to recover the refractive index of the ssDNA layer along and perpendicular to the surface of the 1DPC as well as the vertical growth rate of the ssDNA layer as a function of time. Our approach establishes a technique that can be used for time and polarization-resolved refractometric sensing of bio-analytes and provides a general framework for the monitoring of anisotropic thin films and interfacial processes across both biological and non-biological systems.

\section*{Experimental Measurement}

\begin{figure}[h]
  \centering
  \includegraphics[width=0.5\linewidth]{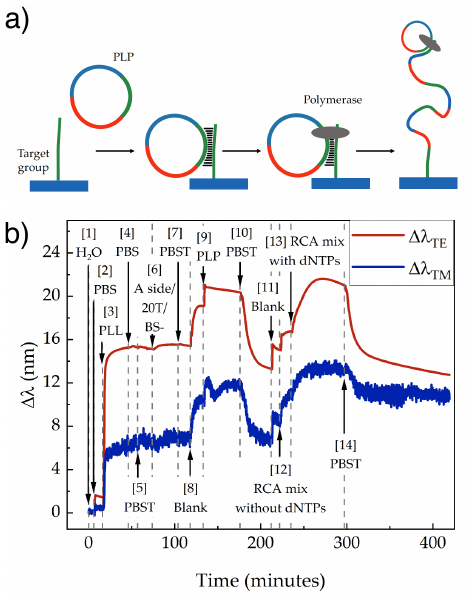}
  \caption{a) Diagram of the Rolling Circle Amplification (RCA) process, showing how the padlock probe (PLP) is used to grow the single-strand DNA (ss-DNA). b) Experimental data of the shift in resonant wavelength for TE and TM BSW modes as a function of time from the growth of the ssDNA via RCA.}
  \label{fig:exp_data}
\end{figure}

To demonstrate the capabilities of our multiplexed BSW refractometric sensing system, we selected DNA as a representative and inherently challenging bio-analyte. DNA is optically complex due to the intrinsic birefringence of its helical structure and its mechanically soft, conformationally dynamic nature. These factors result in nanoscale layers whose refractive index vary based on the grafting density, the orientation of the DNA strands, and the polarization of the probing light. Such characteristics make DNA a stringent test system for evaluating polarization-resolved surface wave sensors. Furthermore, DNA is biologically relevant to many fundamental diseases. In response to these complexities, various nanophotonic platforms have recently emerged to enable the real-time characterization of DNA structures, aiming to provide robust solutions for in vivo diagnostics and cellular research \cite{caixeiroDNASensingWhispering2025,pintoSpectroscopicEllipsometryInvestigation2022,spadavecchiaApproachPlasmonicBased2013,zitoLabelfreeDNABiosensing2021}. Here, we monitor the growth of ssDNA produced through RCA, an isothermal enzymatic process that generates long ssDNA concatemers from a circularized DNA template. When performed on a functionalized substrate, the growing ssDNA remains tethered at one end, forming a brush-like anisotropic layer that grows parallel to the surface normal of the plane on which it is attached with sufficient grafting density. A pictorial diagram of the RCA process is shown in figure \ref{fig:exp_data}a). This dynamic and optically anisotropic system presents a significant challenge for conventional refractometric sensors. By simultaneously tracking the TE and TM BSW resonances, data presented in figure \ref{fig:exp_data}b, our platform provides multiplexed, label-free access to the evolving birefringence during the RCA process.

\begin{figure}[t]
  \centering
  \includegraphics[width=0.4\linewidth]{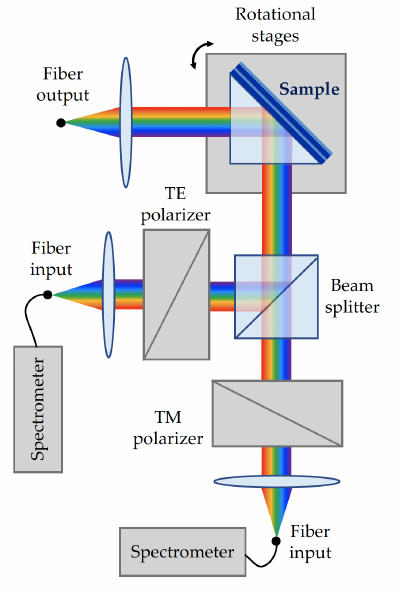}
  \caption{Optical schematic of the multiplexed BSW sensing system.}
  \label{fig:exp_diagram}
\end{figure}

The enabling optical element in our system is a specially designed, truncated one-dimensional photonic crystal (1DPC) capable of simultaneously exciting partially overlapping TE and TM BSW modes \cite{mogniOneDimensionalPhotonicCrystal2022}. Because the dispersion relations of TE and TM BSWs typically differ, their resonant wavelengths diverge rapidly with changes in surface refractive index. This effect can cause one mode to drift outside the detector’s spectral range, particularly during dynamic measurements. To mitigate this, the 1DPC structure was engineered such that the TE and TM dispersion relations remained closely aligned over a broad $(\lambda,\theta_{inc})$ region, minimizing spectral walk-off. This was achieved by properly designing the layer at the top of the multilayer stack, which is known to differentially affect the TE vs TM dispersion relation, allowing for modal overlap. The 1DPC was fabricated using reactive magnetron sputtering and consisted of alternating layers of \ce{SiO2} and \ce{Ta2O5}, capped by a precisely tuned \ce{SiO2} layer. The optical setup used to excite and monitor the BSWs is shown schematically in Fig. \ref{fig:exp_diagram}. Broadband light from a fiber-coupled source was directed onto the 1DPC via a Kretschmann configuration system at an incident angle of $\theta_{\mathrm{inc}} = 66.7$°. The input beam contained both TE and TM components, allowing simultaneous excitation of the two surface modes. The spectra of reflected light was separated into its polarization components using Wollaston prisms and detected by independent spectrometers. The resonant wavelengths of the TE and TM BSW modes were tracked as a function of time during the RCA process. 

The growth of ssDNA via RCA was performed in the aqueous environment which comprised the bulk superstrate for the 1DPC sensing platform. The chemical composition of the aqueous solution was controlled by a microfluidic system designed on the surface of the 1DPC.


 The experimental sequence of the RCA process is described below and the steps appropriately labeled in time in figure ~\ref{fig:exp_data}b. The measurement began by recording the reference spectrum in deionized water (step 1), followed by the introduction of phosphate-buffered saline (PBS, step 2) to stabilize the surface and establish the baseline response in the buffer environment. After 10 minutes of equilibration, a 0.5 mg/mL solution of poly- L-lysine(PLL)- oligo(ethylene glycol) was flowed over the surface for 30 minutes (step 3) to form a physiosorbed polyelectrolyte layer with 6.6$\%$ dibenzocyclooctyne (PLL-OEG-DBCO) serving as functional groups \cite{haslerDualElectronicOptical2025}. The surface was then rinsed with PBS (step 4) to remove unbound molecules. Next, a PBST buffer—PBS supplemented with Tween-20 surfactant—was introduced (step 5) to reduce non-specific adsorption, followed by the injection of the primer sequence (step 6). It is denoted as azide/20T/PS* (40 nM), containing the azide domain for binding to the DBCO-groups on the PLL, a 20-thymine linker region that controls the molecular spacing, and the BS- domain serving as the complementary sequence for subsequent hybridization to the circular padlock probe (PLP). The immobilization of the primer sequence occurred gradually over approximately 25 minutes, during which the resonance shift exhibited a smooth, saturating evolution consistent with surface binding rather than bulk refractive index variations. The surface was again rinsed with PBST (step 7) to remove unbound primers. Subsequently, the buffer solution of the padlock probe was introduced (step 8), producing a distinct refractive index shift associated with the buffer exchange due to the previously performed ex-situ reactions. According to previously published protocols \cite{schmidtRollingCircleAmplification2022}, the linear PLP was circularized in a ligation reaction for 1 hour at 50 °C (containing 75 units of DNA ligase, 0.2 mg/mL bovine serum albumin, 40 nM ligation sequence LS* and ligation buffer), and subsequently treated via incubation with exonuclease I (50 units) and phosphatase (5 units) in the corresponding buffer at 37 °C for 15 minutes to remove all non-circularized DNA with enzymatic inactivation steps in between and at the end for 15 minutes at 85 °C.  After stabilization, the padlock probe PLP was injected for 40 minutes (step 9) to hybridize to the primer sequence BS- on the surface. Both TE and TM resonances displayed clear shifts, indicative of successful hybridization. PBST wash followed (step 10), producing a partial relaxation of the TE mode—likely due to minor baseline drift—while the TM mode remained stable, confirming PLP attachment. To initiate the rolling circle amplification (RCA), the polymerase buffer was introduced (step 11), followed by the addition of 100 units of $\phi$29 DNA polymerase (step 12) and after additional 10 minutes 100 µM of the nucleotide mixtures (step 13). The RCA reaction proceeded isothermally at room temperature for 1 hour, resulting in the continuous extension of DNA strands into brush-like structures on the surface. During this phase, the TE and TM resonances exhibited a pronounced redshift with similar trends to that found in SPR sensing of RCA \cite{xiangRealtimeMonitoringMycobacterium2015,huangProteinDetectionTechnique2007,lechnerSituMonitoringRolling2021}. After an hour, the RCA process was concluded by washing the system with PBST.

\section*{Refractometric Sensitivity Analysis}

To convert the experimentally measured shift in the resonant wavelength, $\lambda_{res}$ as a function of time to a change in refractive index as a function of time, the refractometric sensitivity must be determined. The refractometric sensitivity is generally defined, for our BSW sensor as:

\begin{equation}
    S_{\mathrm{ref}} = \frac{d \lambda_{\mathrm{res}}}{d n}
\end{equation}

\noindent where $n$ is the refractive index of the layer in which the change to the electric permittivity is occurring. In our analysis we will focus on refractive index variations (with respect to pure water) occurring in the superstrate, where the BSW field distribution is present. Specifically in this analysis we will be interested in refractive index dynamics occurring in either the entire superstrate, which we will call the bulk, or to just a thin layer directly above the surface of the 1DPC, which we will denote as the surface layer (SL). The refractometric sensitivity can be determined either experimentally, measuring reflectivity spectra with known refractive index variations, or computationally with a model. In the supplemental section "Experimentally Measured Bulk Calibration", we show an experimentally determined bulk refractometric sensitivity for our sensor. However, for our analysis we will be primarily interested in determining the surface sensitivity ($S_{\mathrm{SL}}$) of a thin layer above the 1DPC surface which requires numerical modeling.

\begin{figure}[h]
  \centering
  \includegraphics[width=0.5\linewidth]{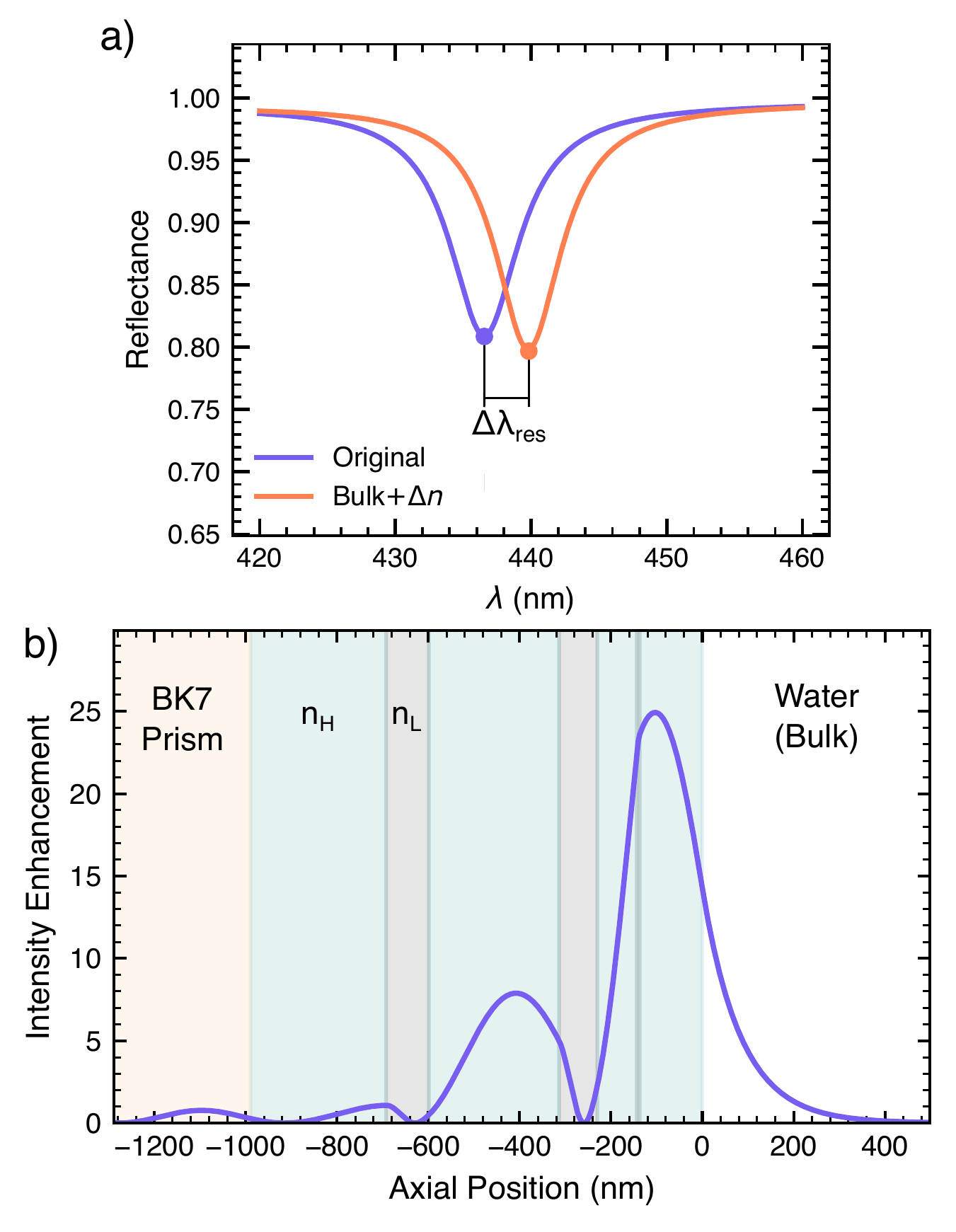}
  \caption{a) Example reflectivity spectra, $R(\lambda)$, for a BSW with and without a perturbation applied to the bulk refractive index of the superstrate. The spectra were calculated using TMM and the wavelength at which these spectra are at a minimum represent the resonant wavelength. In this example, $\Delta n = 20$ mRIU with an associated $\Delta \lambda_{\mathrm{res}} = 3.2$ nm giving a $S_{\mathrm{bulk}} = 324$ $\mathrm{nm}/\mathrm{RIU}$. b) Example field enhancement distribution of our 1DPC without any perturbations applied to our superstrate comprised of an aqueous solution.}
  \label{fig:example_modeling}
\end{figure}

To model the refractometric sensitivity (either $S_{\mathrm{Bulk}}$ or $S_{\mathrm{SL}}$), we used the Transfer Matrix Method (TMM), specifically the implementation (tmmfast) from \cite{luceTMMFastTransferMatrix2022}. Though nominal thicknesses for each layer of the 1DPC were known, fabrication and extinction coefficient uncertainties can significantly alter the photonic band structure of the crystal and thus dramatically change the BSW modes \cite{anopchenkoEffectThicknessDisorder2016a}. To recover more accurate values for the thicknesses of each layer of the 1DPC and the extinction coefficients of the high and low refractive index layers, a genetic optimization was used to minimize the RMS difference between a modeled and experimentally measured reflectivity spectrum, $R(\lambda,\theta_{inc})$, with only an aqueous solution as the bulk superstrate. Details on the fitting procedure are presented in the SI section "Fitting of the 1DPC Parameters".


To calculate the sensitivity, reflectivity spectra were simulated with the TMM before and after a perturbation, $\Delta n$, was applied to the sensing layer, either the bulk superstrate or a thin SL. The resulting shift in the resonant wavelength, $\Delta \lambda_{res}$, was determined by tracking the reflectance minimum ($|r|^2$). An example of this procedure is shown in Fig. \ref{fig:example_modeling}a. The sensitivity was then calculated using a first-order finite difference:

\begin{figure*}[t]
  \centering
  \includegraphics[width=1.0\textwidth]{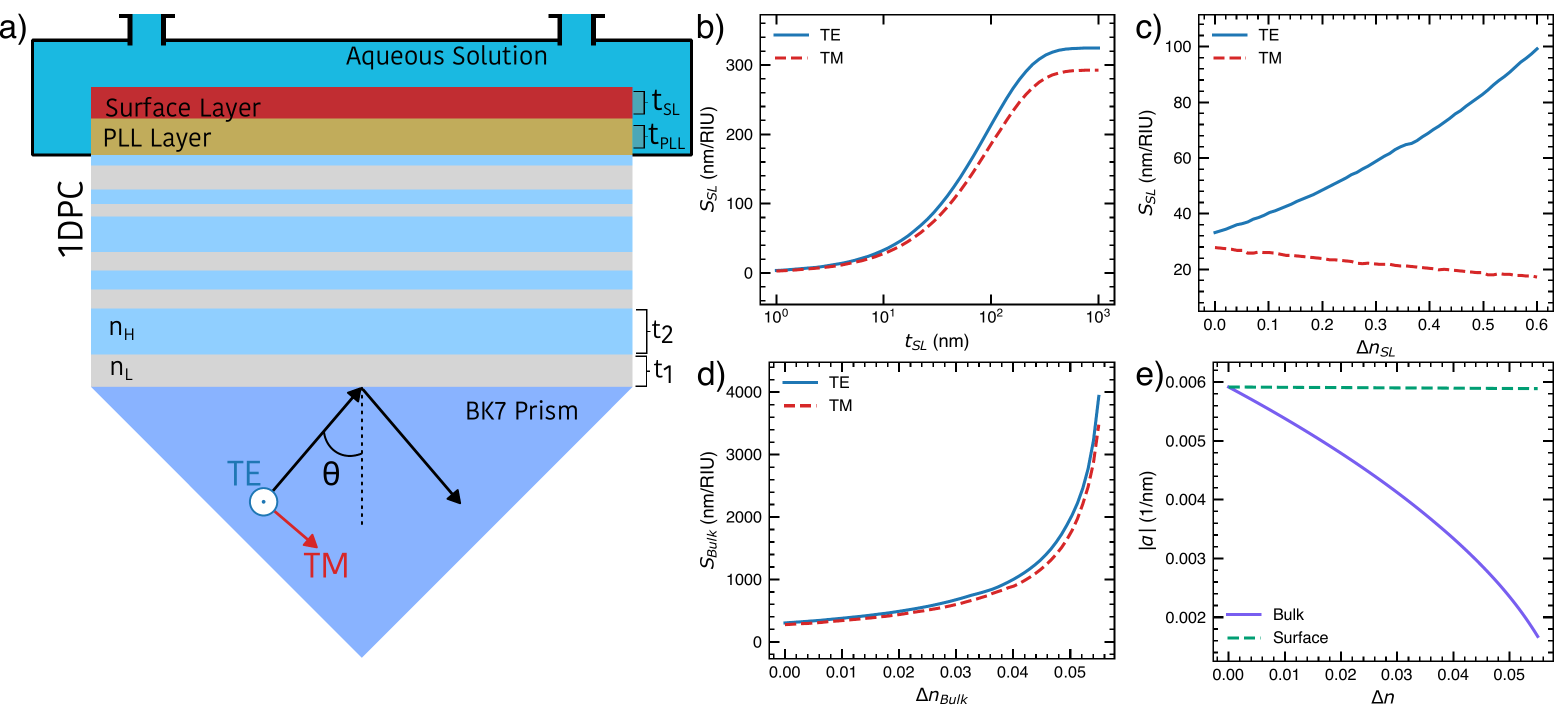}
  \caption{a) Diagram of the 1DPC sensor in the Kretschmann configuration with a 10 nm PLL layer above the 1DPC and a thin Surface Layer, SL, above the PLL layer prior to the aqueous bulk superstrate. b) Refractometric surface layer sensitivity as a function of the thickness of the surface layer. c) Modeled surface sensitivity as a function of the index change applied to the surface layer. d) Modeled bulk sensitivity as a function of the index change applied to the bulk superstrate. e) Exponential decay coefficient, $\alpha$, fitted to the simulated electric field distribution of the exponentially decaying field inside the superstrate as a function of refractive index applied to the bulk and surface layers.}
  \label{fig:surface_sensitivity}
\end{figure*}

\begin{equation}
    S_{bulk,SL} \approx \frac{\Delta \lambda_{res}}{\Delta n_{bulk,SL}} = \frac{min[R_{n+\Delta n}(\lambda)] - min[R_{n}(\lambda)]}{\Delta n_{bulk,SL}}
    \label{eq:modeledSensitivity}
\end{equation}

\noindent where the subscript $n$ of $R_n$ represents the refractive index of the sensing layer. Many biological processes in this study occur within the few nanometers above the 1DPC surface (cite1 stuff?). To accurately model the refractive index variation within this thin layer, we must compute the surface sensitivity which is related to the bulk sensitivity by the fraction of the BSW energy confined within the SL:

\begin{equation}
    S_{\mathrm{SL}} = S_{\mathrm{bulk}}
    \frac{\displaystyle \int_0^{t_{\mathrm{SL}}} 
          |E_{\mathrm{BSW}}(z)|^2   \, dz}
    {\displaystyle \int_0^{\infty} 
          |E_{\mathrm{BSW}}(z)|^2   \, dz},
    \label{eq:surf_sens_full}
\end{equation}

\noindent where $t_{\mathrm{SL}}$ is the surface layer thickness and $E_{\mathrm{BSW}}(z)$ is the axially dependent  electric field of the BSW in the superstrate \cite{diasRefractometricSensitivityBloch2023a}. This equation is validated numerically in SI section "Validation of Relationship Between Surface and Bulk Sensitivity" along with a description of a method for experimental determination of the surface sensitivity.

As shown in Figure \ref{fig:example_modeling}b, the BSW field decays exponentially into the superstrate. This rapid decay makes $S_{\mathrm{SL}}$ highly dependent on the chosen value of $t_{SL}$. Our calculations in Figure \ref{fig:surface_sensitivity}b demonstrate that for a 10 nm SL, the sensitivity is approximately two orders of magnitude lower than the bulk value, a result of the energy distribution within the first few nanometers versus the entire evanescent tail (field distribution shown in figure \ref{fig:example_modeling}b). For $t_{SL} > 500$ nm, the modeled sensitivity converges to the bulk value, reasonably matching the experimental sucrose calibration (see SI Section "Experimentally Measured Bulk Calibration") and validating our numerical approach.

The appropriate value for $t_{SL}$ depends on the step of the experimental sequence, as different processes will occur within different regions of the superstrate. For steps 1 and 2 of the experimental procedure, shown on the data in figure \ref{fig:exp_data}, the chemical changes are largely occurring within the entirety of the bulk superstrate. For step 3, a thin PLL layer is being formed largely within the first 10nm above the 1DPC \cite{haslerDualElectronicOptical2025}. From steps 6 to 13, chemical changes occur within a thin layer above the PLL layer, due to the presence of DNA primer strands and the circular padlock probe. After step 13, the DNA grows and extends into the bulk, which will require a different analysis presented in the section titled "Modeling of DNA Growth". To investigate the surface birefringent dynamics, we assume that chemical changes are occurring only within the 10 nm above the PLL layer, shown as the SL in the diagram in figure \ref{fig:surface_sensitivity} a). The choice of this SL thickness corresponds to an approximate diameter of the circular padlock probe and of the DNA primer strand length. The PLP sequence is estimated to be roughly $~9$ nm in diameter based on the 81 nucleotides in the PLP sequence and the primer strand length is estimated to be around 13 nm based on the 37 nucleotides in the primer strand sequence. Given the complex conformation and the propensity for the hybridized PLP to adopt a tilted orientation relative to the surface, this 10 nm layer effectively encompasses the primary region of refractive index change during the experimental sequence. 

The surface refractometric sensitivity for a 10 nm SL layer calculated in figure \ref{fig:surface_sensitivity}b is calculated with an aqueous superstrate. However, as the refractive index of the surface layer changes due to the chemical processes, the photonic band structure of the 1DPC and electric field distribution of the BSW will change. Therefore, the refractometric sensitivity should be calculated across a range of values of the refractive index of the SL. The calculated $S_{\mathrm{SL}}$ and $S_{\mathrm{Bulk}}$ as a function of refractive index change in the SL/bulk superstrate is shown in figure \ref{fig:surface_sensitivity} c and d respectively. Here we not only see different orders of magnitude for the bulk and surface sensitivities, as expected from the fraction of the BSW field energy contained within the SL, but we also see phenomenologically different behavior of the sensitivity with respect to the change in the refractive index of the sensing layer. In the case of the surface sensitivity, the TE mode $S_{\mathrm{surf}}$ steadily increased with $\Delta n$ meanwhile the TM mode $S_{\mathrm{surf}}$ had the opposite trend. The bulk sensitivity, on the other hand, exhibited an exponential increase of $S_{\mathrm{bulk}}$ for both the TE and TM modes for only small perturbations of $\Delta n_{\mathrm{bulk}}$. The significant difference between these curves can be understood by considering the effect the refractive index change has on the field distribution in the superstrate. The exponential decay of the field distribution in the superstrate is given by:

\begin{equation}
    E_{\mathrm{sup}}(z) = E_{\mathrm{surf}} e^{-\alpha z},
    \label{eq:field_in_sup}
\end{equation}

\noindent where $\alpha = k_0\sqrt{n_{\mathrm{eff}}^2 - n_{\mathrm{sup}}^2}$ and $n_{\mathrm{eff}}$ and $n_{\mathrm{sup}}$ are the effective index of the BSW mode, given by $n_{eff} = n_{prism} \sin{\theta_{res}}$, and the superstrate refractive index. An increase to the bulk refractive index will decrease $\alpha$, causing the field to extend deeper into the bulk superstrate. As a consequence, the percentage of the total BSW energy located inside the superstrate will exponentially increase inducing an exponential increase in the sensitivity. This exponential increase of $\alpha$ will be isotropic for both TE and TM polarization states, which is reflected in the same trends between the two modes in figure \ref{fig:surface_sensitivity}d. When the refractive index is induced in only the SL, the phase matching condition and electric field distribution changes but the decay constant in the superstrate does not significantly vary. Another way to think of this is that changes to the bulk refractive index move the total internal reflection (TIR) angle, meanwhile changes to just the SL do not induce changes to the TIR angle. Figure \ref{fig:surface_sensitivity}e shows fits of $\alpha$ to the calculated electric field distribution in the superstrate as a function of the refractive index change in the sensing layer. These results confirm that the exponential increase of $S_{Bulk}$ occurs due to the increased penetration of the BSW mode into the superstrate. This method presents a way to dramatically increase the sensitivity of any BSW device by increasing the superstrate bulk refractive index without altering the structure of the 1DPC.
\begin{figure*}[t]
  \centering
  \includegraphics[width=1.0\textwidth]{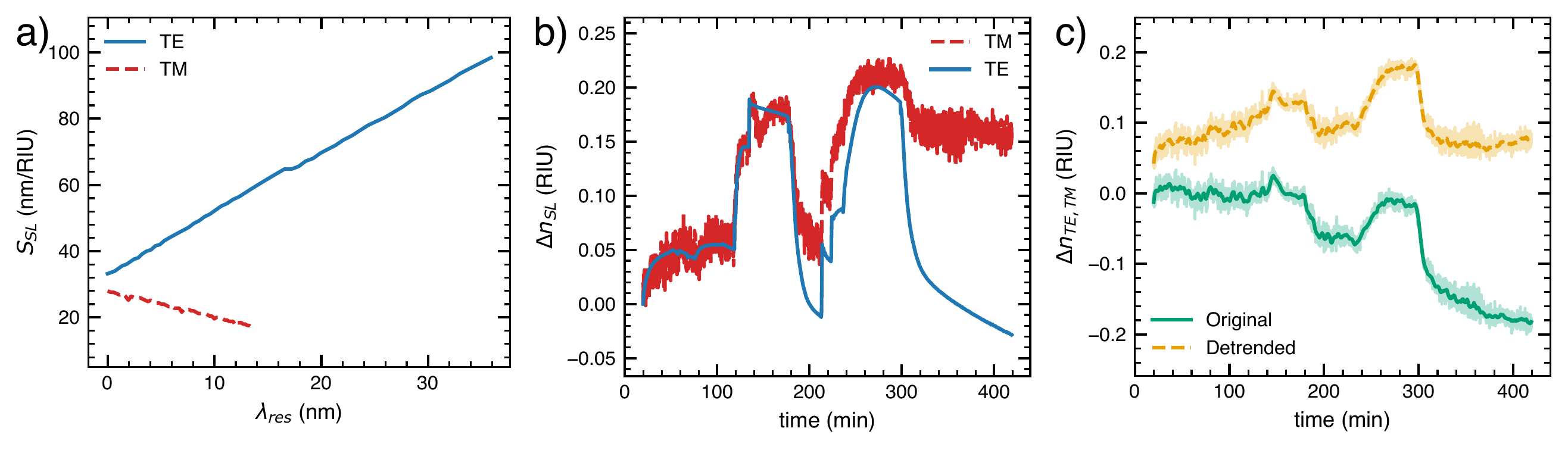}
  \caption{a) Refractometric surface sensitivity as a function of resonant wavelength. b) Experimentally measured resonant wavelength shifts converted to refractive index modifications as a function of time within the surface layer. c) Time dependent birefringence of the surface layer with and without removing the linear phase from the TE data. }
  \label{fig:application_of_wavelength_sensitivity}
\end{figure*}
With the calculated $S_{SL}(\Delta n_{SL})$, we can convert the axes to $S_{SL}(\Delta \lambda_{res})$, plotted explicitly in figure \ref{fig:application_of_wavelength_sensitivity}a, to directly convert the $\lambda_{res}(t)$ to $\Delta n_{SL}$ for both the TE and TM polarization. The sensitivity was calculated with a thin PLL layer at the surface of the 1DPC, therefore the conversion $\Delta \lambda_{res}(t)$ was initiated just after step 3, when the PLL layer formed. The converted experimental data is plotted in figure \ref{fig:application_of_wavelength_sensitivity}b. Large changes to the refractive index of the SL are shown after step 8, when the buffer solution of the PLP was introduced, and after step 13, when the RCA process was initiated. To distinguish between isotropic refractive index variations and birefringent dynamics, we can calculate the birefringence directly through: $\Delta n_{TE,TM}(t) = \Delta n_{SL,TE}(t) - \Delta n_{SL,TM}(t)$. The birefringence as a function of time is plotted in green in figure \ref{fig:application_of_wavelength_sensitivity}c with the lighter line representing the directly converted raw experimental data and the darker line representing the same curve with a Savitzky–Golay filter applied. The birefringence curve shows a relatively constant downward slope through time. This same trend can be clearly seen in the TE resonant wavelength shift as a function of time in figure \ref{fig:exp_data} b) while it cannot be seen in the same curve for the TM polarization state. This constant linear decrease is likely caused from a mechanical drift. The absence of this effect in the TM curve indicates that the drift is unlikely to be caused by slow evaporation of the aqueous solution or by changes in temperature. We can remove this linear trend from the raw data and recover the $\Delta n_{TE,TM}(t)$ curve plotted in yellow as the "detrended" line in figure \ref{fig:application_of_wavelength_sensitivity} c). From this curve we can clearly see that the bulk of the birefringence occurs during the RCA growth process of the ssDNA, while the previous steps, such as the introduction of the PLP and primer strands, results in a significantly smaller birefringence even if the $\Delta n_{SL}(t)$ for the two polarizations are comparable in terms of absolute shift from figure \ref{fig:application_of_wavelength_sensitivity} b). These results show how our polarization multiplexed BSW sensor can be used to distinguish between isotropic refractive index variations and birefringent dynamics. 

\section*{Modeling of DNA Growth}

\begin{figure*}[t]
  \centering
  \includegraphics[width=1.0\textwidth]{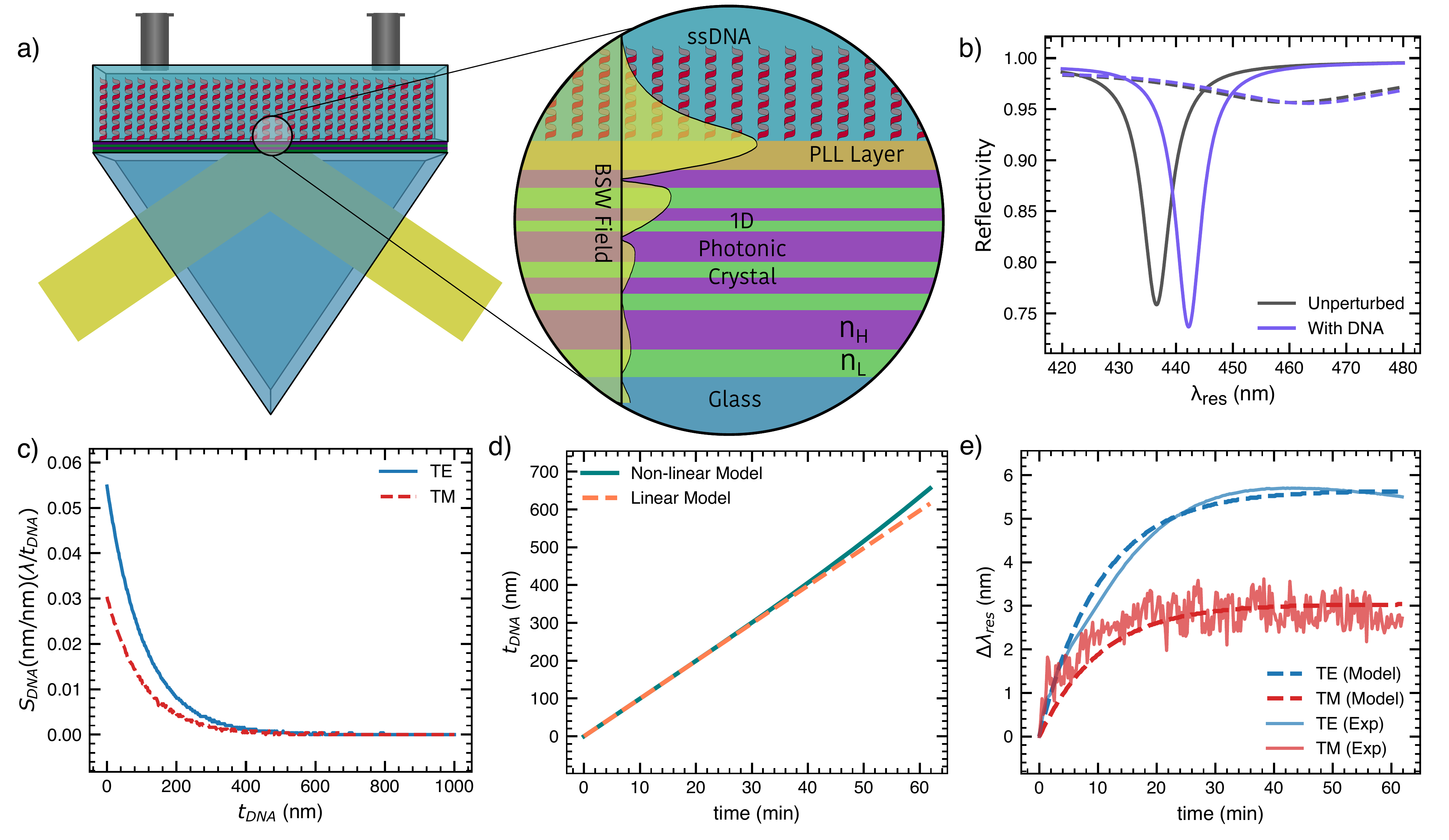}
  \caption{a) Diagram of the BSW sensor with DNA growing on the top layer. b) Reflectivity spectra with and without the DNA present. c) Sensitivity of the DNA layer as a function of the DNA height. d) Thickness as a function of time of the DNA growth during the RCA model fit to the experimentally measured resonant wavelength shifts. The dashed orange line assumed a linear growth rate while the blue solid line allowed for a growth rate up to a polynomial of order 4 with respect to time. e) Fitted model (fitting the growth rate, $n_{TE}$, and $n_{TM}$) to the experimentally measured resonant wavelength shifts during RCA.}
  \label{fig:DNA_growth}
\end{figure*}

Upon the onset of RCA (step 13), the initial model, which assumed refractive index dynamics occurred only within the 10 nm directly above the 1DPC surface, no longer accurately represents the evolving physical system. During RCA, the ssDNA layer is expected to grow in a dense brush which extends axially away from the surface of the 1DPC \cite{lechnerSituMonitoringRolling2021}. The extension of this DNA brush layer is what induces changes to the resonant wavelength, rather than direct refractive index variations within a confined layer above the 1DPC surface. Example TE reflectivity curves with and without a 600 nm ssDNA layer, assumed to have the refractive index found through the analysis in this section, is shown in figure \ref{fig:DNA_growth}b. We model the DNA brush as a variable height, spatially uniform birefringent layer, pictorially represented in figure \ref{fig:DNA_growth}a. The sensitivity of the BSW device can now be defined as $S_{\mathrm{DNA}} = d \lambda/ d t_{\mathrm{DNA}}$ where $t_{\mathrm{DNA}}$ is the axial thickness of the birefringent DNA layer. This thickness sensitivity value will decrease as the ssDNA brush extends into the superstrate, i.e. as $t_{\mathrm{DNA}}$ increases, and will depend on the refractive index values of the birefringent layer along the TE and TM polarizations. A plot of $S_{\mathrm{DNA}}$ as a function of $t_{\mathrm{DNA}}$ is shown in figure \ref{fig:DNA_growth}c and shows the sensitivity asymptotically decreases to 0 as $t_{\mathrm{DNA}}$ extends beyond the axial extent of the BSW field in the superstrate. The differential sensitivity of the TE and TM polarization states is related to the birefringence of the DNA brush, which was modeled according to the refractive indices found in the analysis below. 

 To recover an estimate for the growth rate of the ssDNA brush layer along with the index of refraction of the layer perpendicular (TM) and parallel (TE) to the surface of the 1DPC, we used a nonlinear least square (NLS) optimization method. Specifically, the NLS optimization minimizes the following cost function:

\begin{equation}
    \mathrm{C} = \sqrt{\sum_{\mathrm{TE,TM}}\sum_{t}\left[\lambda_{res,sim}(t,\vec{x_i}) - \lambda_{res,exp}(t)\right]^2},
\end{equation}

\noindent where $t$ is time, $\vec{x_i}$ is the parameter vector for the $i$\textsuperscript{th} iteration of the optimization, and $\lambda_{res,sim}$ is the modeled resonant wavelength shift using the TMM. Within each iteration of the NLS optimization, the reflectivity curve for both the TE and TM polarization states were modeled as a function of time for a given trial parameter set:

\begin{equation}
    \vec{x_i} = \{n_{\mathrm{DNA},\mathrm{TE}},n_{\mathrm{DNA},\mathrm{TM}},\vec{\beta}  \ \}
\end{equation}

\noindent with $\vec{\beta}$ being the coefficients for each polynomial contained in the DNA growth model. The DNA growth model determines the thickness of the birefringent DNA layer as a function of time: $G_{\mathrm{DNA}}(t) = \sum_{j}{\beta_j t^j}$. The resonant wavelengths are calculated from the spectra calculated for each time/thickness value and the cost function is evaluated at each step of the optimization. Optimizations were performed assuming both a strictly linear growth model and a nonlinear growth model with up to the 4 \textsuperscript{th} order time polynomial. The resulting $\lambda_{res,sim}(t)$, for the linear DNA growth model, is plotted for both the TE and TM polarization states on top of the $\lambda_{res,exp}(t)$ in figure \ref{fig:DNA_growth} e showing strong agreement. The retrieved refractive index of the DNA layer along the direction of the electric field for the TE polarized light was $n_{DNA,TE} = 1.356$ and $n_{DNA,TM} = 1.349$ along the electric field of the TM polarization. The exact refractive index will strongly depend on the brush density and conformation of the DNA layer, however the determined birefringence of 0.006 RIU falls within the range previously reported in the literature \cite{lechnerSituMonitoringRolling2021,caixeiroDNASensingWhispering2025,samocRefractiveindexAnisotropyOptical2007}. With the linear growth model, the NLS optimization retrieved a final DNA length of 620 nm which falls in line with the expected range given the drop off in sensitivity. Expected literature values for the growth rate vary greatly depending on the experimental conditions, and given the likely complex conformation of the ssDNA strands, comparison to previously reported literature values would be difficult. Lechner et al \cite{lechnerSituMonitoringRolling2021} , however presented growth rates on the same order of magnitude with similar RCA procedures. The RCA process should have a linear growth rate as discussed in the literature \cite{nallurSignalAmplificationRolling2001}. We can test the linearity of the growth rate by investigating the results from a NLS optimization with a nonlinear growth rate model. The retrieved growth rate from this optimization is plotted in figure \ref{fig:DNA_growth}d in turquoise, and exhibits an almost perfectly linear growth rate. This result is a good confirmation that we are accurately modeling the growth of the DNA brush with our birefringent slab model. The small differences between the model and the experimentally measured resonant wavelength shifts are likely due to conformation changes during the process or other experimental factors such as temperature fluctuations. 

\section*{Conclusion}
In this work, we demonstrated a polarization-multiplexed refractometric sensing platform based on BSWs in 1DPCs, capable of simultaneous TE and TM mode excitation. By monitoring resonant wavelength shifts for both polarizations in real-time, we directly probed the birefringent dynamics of a biointerface. This platform was applied to track the growth of ssDNA during RCA. To accurately analyze this process, we implemented a two-stage modeling approach based on the Transfer Matrix Method. In the first stage, we utilized a wavelength-dependent surface sensitivity model that confined refractive index changes to a thin layer (~10 nm) directly above the crystal surface. This allowed us to distinguish isotropic background variations during initial surface stabilization from true birefringent signals. In the second stage, following the onset of RCA, we transitioned to a birefringent slab model to account for the vertical extension of the DNA brush into the superstrate. By applying a nonlinear least-squares optimization to this growth model, we recovered a final ssDNA layer thickness of approximately 620 nm and a persistent birefringence of 0.006 RIU ($n_{\mathrm{DNA,TE}}=1.356$ and $n_{\mathrm{DNA,TM}}=1.349$). These results highlight the platform’s capacity for label-free, time-resolved characterization of complex, anisotropic biomolecular interfaces, providing a robust framework for investigating dynamic structural evolutions in biological thin films.
\section{Supplemental Information}

\subsection{Fitting of the 1DPC Parameters}

\begin{figure}[h!]
  \centering
  \includegraphics[width=0.5\linewidth]{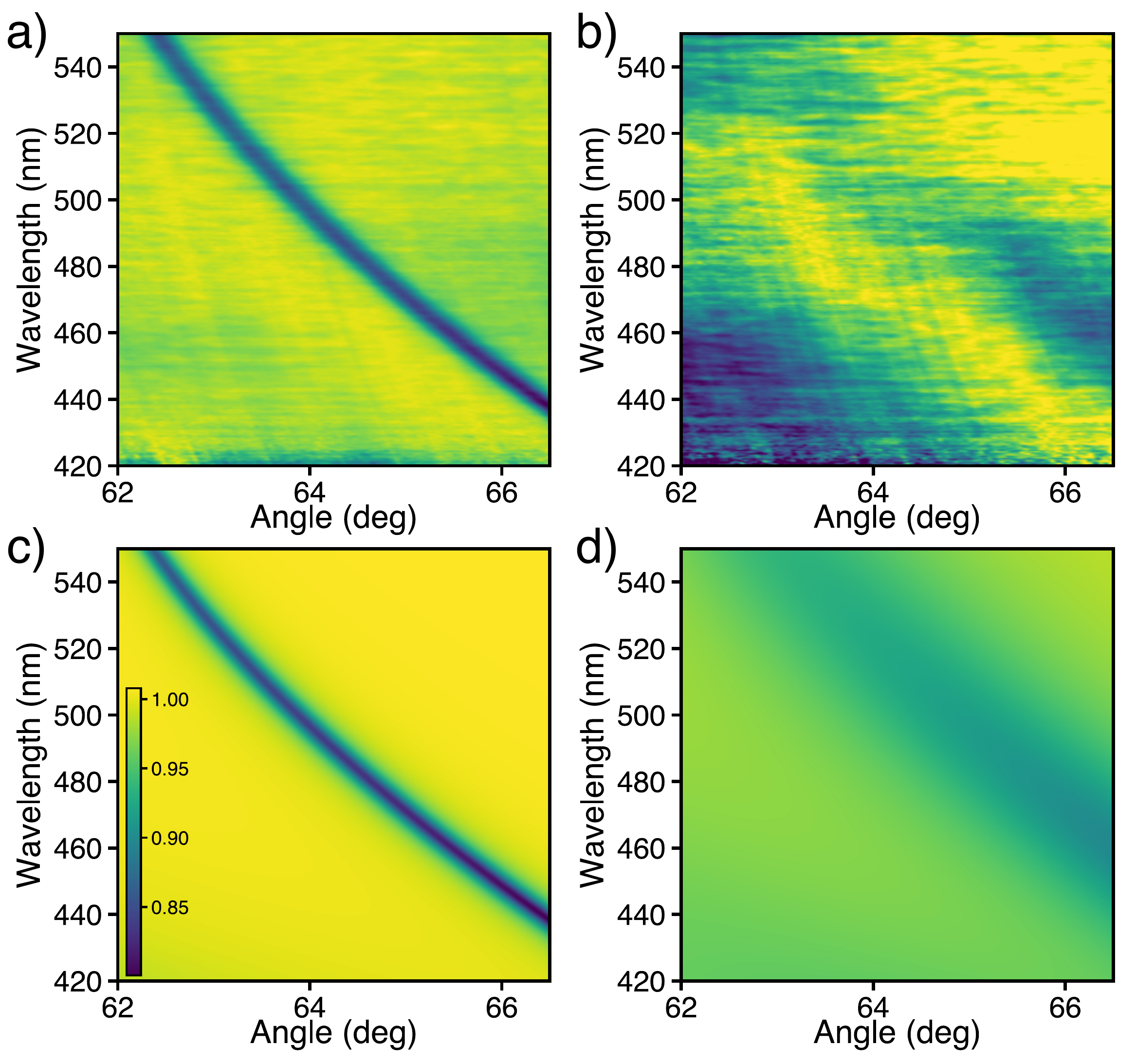}
  \caption{a) and b) show the experimentally measured reflectivity maps in $(\lambda,\theta_{inc})$ space for the 1DPC with an aqueous solution as the superstrate. c) and d) show the simulated reflectivity maps for the 1DPC whose structure was genetically minimized to have the comparable reflectivity maps to the experimentally measured reflectivity maps.}
  \label{fig:1DPC_modeling}
\end{figure}

To recover more accurate values for the layers of the 1DPC and the extinction coefficients of the high and low refractive index materials, an optimization routine was employed to compare modeled reflectivity spectra to the experimentally measured reflectivity spectra. The experimental reflectivity spectra was measured with an aqueous superstrate along in a $(\lambda,\theta_{inc})$ region over which the BSW could be excited, plotted in figure \ref{fig:1DPC_modeling}a and b. Using the TMM method, a reflectivity spectra in the same $(\lambda,\theta_{inc})$ region was numerically calculated for a given set of trial parameters:

\begin{equation}
    x_{\mathrm{param}}: \{t_{1},t_{2},...,t_{N},k_{\ce{SiO2}},k_{\ce{Ta2O5}}\},
\end{equation}
\noindent where $t_n$ is the thickness of the $n^{\text{th}}$ layer and $k$ is the extinction coefficient of the two materials forming the high- and low-index layers of the 1DPC. The parameters were optimized using a genetic optimization to minimize the following evaluation function:

\begin{equation}
    \mathrm{Eval} = \sum_{\mathrm{TE,TM}}\frac{1}{2}\sqrt{\sum_{\lambda} \sum_{\theta_{\mathrm{inc}}}\left[R_{sim}(\lambda,\theta_{\mathrm{inc}}) - R_{exp}(\lambda,\theta_{\mathrm{inc}})\right]^2},
\end{equation}
\noindent which computes the root-mean-square error across both polarizations and all measured wavelengths and incident angles. The optimization ran for 300 generations with 300 individuals per generation and a crossover probability of 0.25. The best individual achieved an evaluation value of 0.019, indicating good agreement between the simulated and experimental reflectivity maps. The simulated reflectivity maps for the optimized structure are shown in Fig.~\ref{fig:1DPC_modeling} c),d) for TE and TM polarizations, respectively. The nominal and recovered thicknesses for the 

\begin{table}[h]
\centering
\begin{tabular}{@{}ccc@{}}
\toprule
\textbf{Layer} & \textbf{Recovered Thicknesses (nm)} \\
\midrule
1   & Substrate     \\
2   & 297        \\
3   & 94        \\
4   & 285        \\
5    & 84        \\
6   & 85        \\
7     & 7        \\
8  & 137        \\
9           & Superstrate   \\
\bottomrule
\end{tabular}
\caption{Recovered layer thicknesses from the genetic minimization of the RMS difference between the modeled and experimentally measured reflectivity spectra.}
\label{tab:layer_thicknesses}
\end{table}

\subsection{Experimentally Measured Bulk Calibration}

\begin{figure*}[t]
  \centering
  \includegraphics[width=1.0\textwidth]{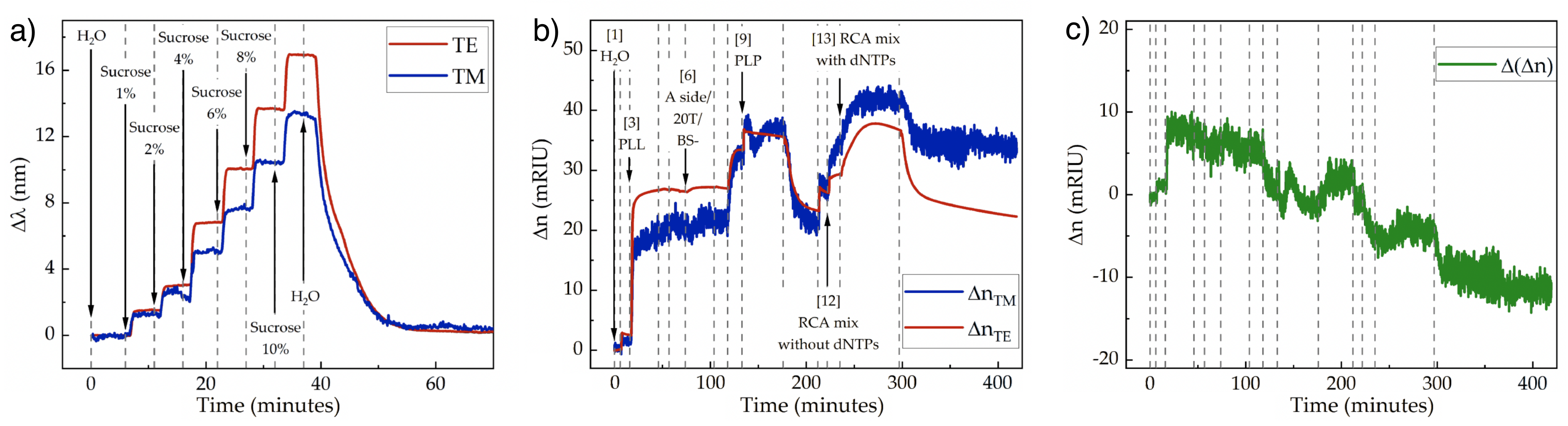}
  \caption{a) Experimental results from a bulk sucrose calibration method used to determine the sensitivity of our 1DPCs refractometric sensing device to changes in the index of refraction at the top surface. b) Resulting application of the sensitivity obtained to the time evolution of the resonant wavelength of the BSW mode for TE and TM polarizations. c) time evolution of the birefringence of the system calculated from the application of the bulk sucrose calibration. }
  \label{fig:bulk_sensitivity}
\end{figure*}

To compliment the numerically calculated bulk sensitivity values for our BSW sensing device, a sucrose calibration test was performed to experimentally determine our bulk refractometric sensitivity. The sucrose calibration was done by taking reflectivity spectra while sucrose was dissolved within the aqueous solution in known quantities. The shift of the resonant wavelength was recorded and correlated to the shift in the refractive index caused by the dissolved sucrose. The resonant wavelength shifts as a function of time for each step of the sucrose calibration is shown in figure \ref{fig:bulk_sensitivity}a. The experimentally recovered sensitivities are 595 nm/RIU and 340 nm/RIU for the TE and TM BSW modes respectively, and exhibit the same increase in sensitivity at higher sucrose calibrations as is expected from the analysis in the paper. The sensitivities from the bulk sucrose calibration are used to calculate the refractive index change of the bulk superstrate layer in time for the TE and TM polarization states, shown in figure \ref{fig:bulk_sensitivity}b. Subtracting the two curves gives us the birefringence as a function of time, which strongly differs from the results obtained with our surface sensitivity analysis. This result shows the importance of carefully determining the refractometric sensitivity of the 1DPC device when recovering birefringence curves. 

\section{DNA Sequences and PLL Chemistry}

\begin{figure}[t]
  \centering
  \includegraphics[width=0.5\linewidth]{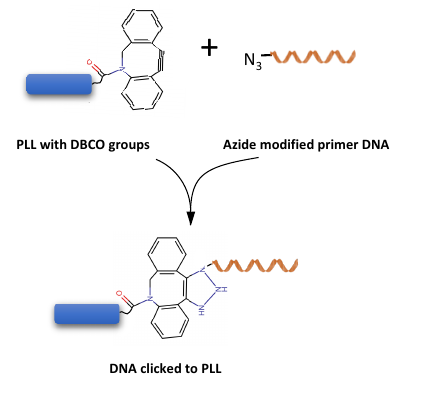}
  \caption{Schematical drawing for the click-chemistry reaction to link the DNA primer to the PLL-OEG-DBCO surface.}
  \label{fig:clickChemisty}
\end{figure}

\begin{table*}[h]
\centering
\renewcommand{\arraystretch}{1.5}
\begin{tabularx}{\textwidth}{l X}
\toprule
\textbf{DNA probe} & \textbf{Sequence 5' $\rightarrow$ 3'} \\ \midrule
Linear padlock probe (PLP) & \texttt{\textcolor{green!50!black}{TG TGA TAC AGC TTT CTT} GCGC GTG TAT GCA GCT CCT CGA GTA GC \textcolor{red!80!black}{C GCA GTT CGC GCC GCA G} GG \textcolor{green!50!black}{CCG ATA CGT GTA ACT TAT}} \\ \midrule
Ligation sequence (LS*) & \texttt{AA\textcolor{green!50!black}{G AAA GCT GTA TCA CA ATA AGT TAC ACG TAT CGG}} \\ \midrule
Primer sequence (azide/20T/PS*) & \texttt{azide-TTTTTTTTTTTTTTTTTTTT\textcolor{red!80!black}{CTGCGGCGCGAACTGCG}} \\ \bottomrule
\end{tabularx}
\caption{DNA sequences for the RCA process – the colors indicate the complementary sequences to the padlock probe, marked with (*).}
\label{table:DNASequence}
\end{table*}

The DNA sequences for the primer strands, circular padlock probe (PLP), and ligation sequence (LS*) are summarized in Table \ref{table:DNASequence}. In this table, color-coding is used to indicate the complementary domains between the different sequences, highlighting the hybridization sites necessary for successful ligation and rolling circle amplification.

To facilitate the covalent attachment of the primer DNA to the 1DPC surface, a functionalized copolymer, poly-L-lysine-oligo(ethylene glycol) modified with dibenzocyclooctyne (PLL-g-OEG-DBCO), was employed as an interfacial layer. A schematic representation of the click reaction between the azide-terminated DNA primer and the DBCO-functionalized surface is illustrated in Figure \ref{fig:clickChemisty}.

\subsection{Validation of Relationship Between Surface and Bulk Sensitivity}

\begin{figure}[t]
  \centering
  \includegraphics[width=0.5\linewidth]{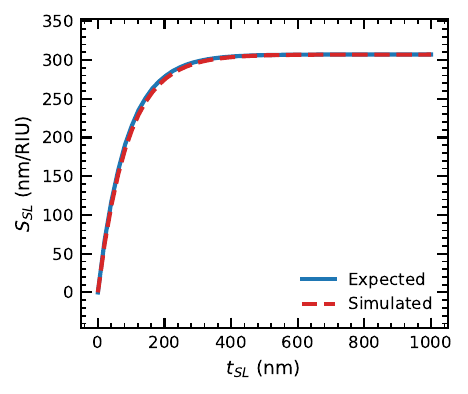}
  \caption{Plot of the surface sensitivity as a function of thickness of the surface layer. The solid blue curve is calculated from equation \ref{eq:surf_sens_full} while the red solid curve is calculated numerically from simulated reflectivity spectra. }
  \label{fig:validation_surface_sensitivity}
\end{figure}

The equation relating the surface sensitivity to the bulk sensitivity in the main text, equation \ref{eq:surf_sens_full}, can be derived using a perturbation approach similar to that prescribed in \cite{diasRefractometricSensitivityBloch2023a}. The refractometric sensitivity is defined as:

\begin{equation}
    S_{ref} = \frac{\partial \lambda_{res}}{\partial n} = \frac{\partial \lambda_{res}}{\partial n_{eff}} \frac{\partial n_{eff}}{\partial n},
\end{equation}

\noindent where $n_{eff}$ is the effective index of the BSW mode and $n$ is the refractive index of the sensing layer (either the bulk or the SL). The $\partial \lambda_{res}/\partial n_{eff}$ is explained in reference \cite{diasRefractometricSensitivityBloch2023a} and the instrumental contribution to the sensitivity which depends on the method in which the BSW is excited. For prism excitation, like in our device, the instrumental contribution will depend on the chromatic dispersion of the prism and the effective index of the BSW. 

To calculate the second term, we can use the same perturbation theory approach as \cite{diasRefractometricSensitivityBloch2023a}. Below we will show the calculation for the TE mode, however the TM mode follows a similar form. A small perturbation to the effective index of the BSW can be calculated as:

\begin{equation}
    \Delta(n^2_{eff}) = \frac{\int^{\infty}_{-\infty}\Delta\epsilon(z)|E(z)|^2dz}{\int^{\infty}_{-\infty}|E(z)|^2dz}
\end{equation}

\noindent where $\Delta \epsilon(z)=2 n \Delta{n}$ which is the refractive index and change of the refractive index of the sensing layer. Inserting the perturbation to the axially dependent electric permittivity:

\begin{equation}
    \Delta(n^2_{eff}) = 2 n_{eff} \Delta n_{eff} = 2 n \Delta n \frac{\int^{t_{SL}}_{0}|E(z)|^2dz}{\int^{\infty}_{-\infty}|E(z)|^2dz}
\end{equation}

\noindent which can be rearranged to get:

\begin{equation}
    \frac{\partial n_{eff}}{\partial n} \approx \frac{\Delta n_{eff}}{\Delta n} = \frac{n}{n_{eff}} \frac{\int^{t_{SL}}_{0}|E(z)|^2dz}{\int^{\infty}_{-\infty}|E(z)|^2dz}.
\end{equation}

The refractometric sensitivity for perturbations to the refractive index of a thin surface sensing layer versus perturbations applied to the whole bulk, only differ by $t_{SL}$ which is finite for a surface sensitivity and infinite for the bulk. With this we recover the relationship between the surface sensitivity and the bulk sensitivity:

\begin{equation}
    S_{\mathrm{surface}} = S_{\mathrm{bulk}}
    \frac{\displaystyle \int_0^{t_{\mathrm{SL}}} 
          |E_{\mathrm{BSW}}(z)|^2   \, dz}
    {\displaystyle \int_0^{\infty} 
          |E_{\mathrm{BSW}}(z)|^2   \, dz}.   
\end{equation}

\noindent Given our equation for the evanescent field in the superstrate, we recover:

\begin{equation}
    S_{\mathrm{surface}} = S_{\mathrm{bulk}} (1-e^{-2 \alpha t_{SL}}),
    \label{eq:surf_sens_simple}
\end{equation}

\noindent with $\alpha$ being the exponential decay of the field in the superstrate. We can further verify this relationship numerically by comparing this result to sensitivities calculated directly from reflectivity spectra. First, we calculated the expected results from equation \ref{eq:surf_sens_full} by simulating the bulk refractometric sensitivity and the electric field in the superstrate as a function of distance within the bulk superstrate. The resulting surface sensitivity from this calculation is plotted in blue in figure \ref{fig:validation_surface_sensitivity}. A simulated surface sensitivity is then calculated as a function of thickness of the surface layer, as is done in the main manuscript by calculating reflectivity spectra for various perturbations of the surface layer refractive index. The results from the simulated surface sensitivity are plotted with a dashed red line in figure \ref{fig:validation_surface_sensitivity}. The two results agree showing almost perfect overlap of the curves, verifying equation \ref{eq:surf_sens_full} in the main text. While here the electric field distributions are calculated using a TMM method, equation \ref{eq:field_in_sup} could be used in conjunction with an experimentally measured bulk calibration to calculate the surface sensitivity experimentally. 

\bibliographystyle{IEEEtran}
\bibliography{DNASensingbib}

@article{arrudaUltrasensitiveAlzheimersDisease2025,
  title = {Ultrasensitive {{Alzheimer}}'s Disease Biomarker Detection with Nanopillar Photonic Crystal Biosensors},
  author = {Arruda, Guilherme S. and Morris, Katie and Martins, Augusto and Wang, Yue and {Sloan-Dennison}, Sian and Graham, Duncan and Quinn, Steven D. and Martins, Emiliano R. and Krauss, Thomas F.},
  year = 2025,
  month = oct,
  journal = {Optica},
  volume = {12},
  number = {10},
  pages = {1587},
  issn = {2334-2536},
  doi = {10.1364/OPTICA.566672},
  urldate = {2025-10-10},
  langid = {english}
}

@article{aurelioElectromagneticFieldEnhancement2017a,
  title = {Electromagnetic Field Enhancement in {{Bloch}} Surface Waves},
  author = {Aurelio, Daniele and Liscidini, Marco},
  year = 2017,
  month = jul,
  journal = {Physical Review B},
  volume = {96},
  number = {4},
  pages = {045308},
  issn = {2469-9950, 2469-9969},
  doi = {10.1103/PhysRevB.96.045308},
  urldate = {2025-02-11},
  copyright = {http://link.aps.org/licenses/aps-default-license},
  langid = {english}
}

@article{caixeiroDNASensingWhispering2025,
  title = {{{DNA Sensing}} with {{Whispering Gallery Mode Microlasers}}},
  author = {Caixeiro, Soraya and D{\"o}rrenhaus, Robert and Popczyk, Anna and Schubert, Marcel and {Kath-Schorr}, Stephanie and Gather, Malte C.},
  year = 2025,
  month = mar,
  journal = {Nano Letters},
  volume = {25},
  number = {11},
  pages = {4467--4475},
  issn = {1530-6984, 1530-6992},
  doi = {10.1021/acs.nanolett.5c00078},
  urldate = {2025-10-08},
  copyright = {https://creativecommons.org/licenses/by/4.0/},
  langid = {english}
}

@article{descroviCouplingSurfaceWaves2007,
  title = {Coupling of Surface Waves in Highly Defined One-Dimensional Porous Silicon Photonic Crystals for Gas Sensing Applications},
  author = {Descrovi, Emiliano and Frascella, Francesca and Sciacca, Beniamino and Geobaldo, Francesco and Dominici, Lorenzo and Michelotti, Francesco},
  year = 2007,
  month = dec,
  journal = {Applied Physics Letters},
  volume = {91},
  number = {24},
  pages = {241109},
  issn = {0003-6951, 1077-3118},
  doi = {10.1063/1.2824387},
  urldate = {2025-10-28},
  langid = {english}
}

@article{diasRefractometricSensitivityBloch2023a,
  title = {Refractometric Sensitivity of {{Bloch}} Surface Waves: Perturbation Theory Calculation and Experimental Validation},
  shorttitle = {Refractometric Sensitivity of {{Bloch}} Surface Waves},
  author = {Dias, Bernardo Santos and De Almeida, Jos{\'e} M. M. M. and Coelho, Lu{\'i}s C. C.},
  year = 2023,
  month = feb,
  journal = {Optics Letters},
  volume = {48},
  number = {3},
  pages = {727},
  issn = {0146-9592, 1539-4794},
  doi = {10.1364/OL.481176},
  urldate = {2025-10-07},
  langid = {english}
}

@article{grygaBlochSurfaceWave2020,
  title = {Bloch {{Surface Wave Resonance Based Sensors}} as an {{Alternative}} to {{Surface Plasmon Resonance Sensors}}},
  author = {Gryga, Michal and Ciprian, Dalibor and Hlubina, Petr},
  year = 2020,
  month = sep,
  journal = {Sensors},
  volume = {20},
  number = {18},
  pages = {5119},
  issn = {1424-8220},
  doi = {10.3390/s20185119},
  urldate = {2025-10-08},
  copyright = {https://creativecommons.org/licenses/by/4.0/},
  langid = {english}
}

@article{homolaSurfacePlasmonResonance2008,
  title = {Surface {{Plasmon Resonance Sensors}} for {{Detection}} of {{Chemical}} and {{Biological Species}}},
  author = {Homola, Ji{\v r}{\'i}},
  year = 2008,
  month = feb,
  journal = {Chemical Reviews},
  volume = {108},
  number = {2},
  pages = {462--493},
  issn = {0009-2665, 1520-6890},
  doi = {10.1021/cr068107d},
  urldate = {2025-10-08},
  langid = {english}
}

@article{kabashinPhaseAmplitudeSensitivities2009,
  title = {Phase and Amplitude Sensitivities in Surface Plasmon Resonance Bio and Chemical Sensing},
  author = {Kabashin, Andrei V. and Patskovsky, Sergiy and Grigorenko, Alexander N.},
  year = 2009,
  month = nov,
  journal = {Optics Express},
  volume = {17},
  number = {23},
  pages = {21191},
  issn = {1094-4087},
  doi = {10.1364/OE.17.021191},
  urldate = {2025-10-28},
  copyright = {https://doi.org/10.1364/OA\_License\_v1\#VOR-OA},
  langid = {english}
}

@article{khanBlochSurfaceWave2016,
  title = {Bloch Surface Wave Structures for High Sensitivity Detection and Compact Waveguiding},
  author = {Khan, Muhammad Umar and Corbett, Brian},
  year = 2016,
  month = jan,
  journal = {Science and Technology of Advanced Materials},
  volume = {17},
  number = {1},
  pages = {398--409},
  issn = {1468-6996, 1878-5514},
  doi = {10.1080/14686996.2016.1202082},
  urldate = {2025-10-07},
  langid = {english}
}

@article{lereuSurfacePlasmonsBloch2017,
  title = {Surface Plasmons and {{Bloch}} Surface Waves: {{Towards}} Optimized Ultra-Sensitive Optical Sensors},
  shorttitle = {Surface Plasmons and {{Bloch}} Surface Waves},
  author = {Lereu, A. L. and Zerrad, M. and Passian, A. and Amra, C.},
  year = 2017,
  month = jul,
  journal = {Applied Physics Letters},
  volume = {111},
  number = {1},
  pages = {011107},
  issn = {0003-6951, 1077-3118},
  doi = {10.1063/1.4991358},
  urldate = {2025-10-08},
  langid = {english}
}

@article{mogniOneDimensionalPhotonicCrystal2022,
  title = {One-{{Dimensional Photonic Crystal}} for {{Surface Mode Polarization Control}}},
  author = {Mogni, Erika and Pellegrini, Giovanni and Gil-Rostra, Jorge and Yubero, Francisco and Simone, Giuseppina and Fossati, Stefan and Dost{\'a}lek, Jakub and Mart{\'i}nez V{\'a}zquez, Rebeca and Osellame, Roberto and Celebrano, Michele and Finazzi, Marco and Biagioni, Paolo},
  year = 2022,
  month = nov,
  journal = {Advanced Optical Materials},
  volume = {10},
  number = {21},
  pages = {2200759},
  issn = {2195-1071, 2195-1071},
  doi = {10.1002/adom.202200759},
  urldate = {2024-03-04},
  langid = {english}
}

@article{occhiconeSpectralCharacterizationMidInfrared2021a,
  title = {Spectral {{Characterization}} of {{Mid-Infrared Bloch Surface Waves Excited}} on a {{Truncated 1D Photonic Crystal}}},
  author = {Occhicone, Agostino and Pea, Marialilia and Polito, Raffaella and Giliberti, Valeria and Sinibaldi, Alberto and Mattioli, Francesco and Cibella, Sara and Notargiacomo, Andrea and Nucara, Alessandro and Biagioni, Paolo and Michelotti, Francesco and Ortolani, Michele and Baldassarre, Leonetta},
  year = 2021,
  month = jan,
  journal = {ACS Photonics},
  volume = {8},
  number = {1},
  pages = {350--359},
  issn = {2330-4022, 2330-4022},
  doi = {10.1021/acsphotonics.0c01657},
  urldate = {2025-10-07},
  copyright = {http://pubs.acs.org/page/policy/authorchoice\_ccby\_termsofuse.html},
  langid = {english}
}

@article{pintoSpectroscopicEllipsometryInvestigation2022,
  title = {Spectroscopic {{Ellipsometry Investigation}} of a {{Sensing Functional Interface}}: {{DNA SAMs Hybridization}}},
  shorttitle = {Spectroscopic {{Ellipsometry Investigation}} of a {{Sensing Functional Interface}}},
  author = {Pinto, Giulia and Dante, Silvia and Rotondi, Silvia Maria Cristina and Canepa, Paolo and Cavalleri, Ornella and Canepa, Maurizio},
  year = 2022,
  month = jul,
  journal = {Advanced Materials Interfaces},
  volume = {9},
  number = {19},
  pages = {2200364},
  issn = {2196-7350, 2196-7350},
  doi = {10.1002/admi.202200364},
  urldate = {2025-10-08},
  langid = {english}
}

@article{rizzoBlochSurfaceWave2018,
  title = {Bloch Surface Wave Label-Free and Fluorescence Platform for the Detection of {{VEGF}} Biomarker in Biological Matrices},
  author = {Rizzo, Riccardo and Alvaro, Maria and Danz, Norbert and Napione, Lucia and Descrovi, Emiliano and Schmieder, Stefan and Sinibaldi, Alberto and Chandrawati, Rona and Rana, Subinoy and Munzert, Peter and Schubert, Thomas and Maillart, Emmanuel and Anopchenko, Aleksei and Rivolo, Paola and Mascioletti, Alessandro and Sonntag, Frank and Stevens, Molly M. and Bussolino, Federico and Michelotti, Francesco},
  year = 2018,
  month = feb,
  journal = {Sensors and Actuators B: Chemical},
  volume = {255},
  pages = {2143--2150},
  issn = {09254005},
  doi = {10.1016/j.snb.2017.09.018},
  urldate = {2025-10-07},
  langid = {english}
}

@article{sinibaldiBlochSurfaceWaves2017,
  title = {Bloch {{Surface Waves Biosensors}} for {{High Sensitivity Detection}} of {{Soluble ERBB2}} in a {{Complex Biological Environment}}},
  author = {Sinibaldi, Alberto and Sampaoli, Camilla and Danz, Norbert and Munzert, Peter and Sonntag, Frank and Centola, Fabio and Occhicone, Agostino and Tremante, Elisa and Giacomini, Patrizio and Michelotti, Francesco},
  year = 2017,
  month = aug,
  journal = {Biosensors},
  volume = {7},
  number = {3},
  pages = {33},
  issn = {2079-6374},
  doi = {10.3390/bios7030033},
  urldate = {2025-10-07},
  copyright = {https://creativecommons.org/licenses/by/4.0/},
  langid = {english}
}

@article{sinibaldiDirectComparisonPerformance2012,
  title = {Direct Comparison of the Performance of {{Bloch}} Surface Wave and Surface Plasmon Polariton Sensors},
  author = {Sinibaldi, Alberto and Danz, Norbert and Descrovi, Emiliano and Munzert, Peter and Schulz, Ulrike and Sonntag, Frank and Dominici, Lorenzo and Michelotti, Francesco},
  year = 2012,
  month = nov,
  journal = {Sensors and Actuators B: Chemical},
  volume = {174},
  pages = {292--298},
  issn = {09254005},
  doi = {10.1016/j.snb.2012.07.015},
  urldate = {2025-10-28},
  copyright = {https://www.elsevier.com/tdm/userlicense/1.0/},
  langid = {english}
}

@article{spadavecchiaApproachPlasmonicBased2013,
  title = {Approach for {{Plasmonic Based DNA Sensing}}: {{Amplification}} of the {{Wavelength Shift}} and {{Simultaneous Detection}} of the {{Plasmon Modes}} of {{Gold Nanostructures}}},
  shorttitle = {Approach for {{Plasmonic Based DNA Sensing}}},
  author = {Spadavecchia, Jolanda and Barras, Alexandre and Lyskawa, Joel and Woisel, Patrice and Laure, William and Pradier, Claire-Marie and Boukherroub, Rabah and Szunerits, Sabine},
  year = 2013,
  month = mar,
  journal = {Analytical Chemistry},
  volume = {85},
  number = {6},
  pages = {3288--3296},
  issn = {0003-2700, 1520-6882},
  doi = {10.1021/ac3036316},
  urldate = {2025-10-08},
  langid = {english}
}

@article{steglichSurfacePlasmonResonance2022,
  title = {Surface {{Plasmon Resonance}} ({{SPR}}) {{Spectroscopy}} and {{Photonic Integrated Circuit}} ({{PIC}}) {{Biosensors}}: {{A Comparative Review}}},
  shorttitle = {Surface {{Plasmon Resonance}} ({{SPR}}) {{Spectroscopy}} and {{Photonic Integrated Circuit}} ({{PIC}}) {{Biosensors}}},
  author = {Steglich, Patrick and Lecci, Giulia and Mai, Andreas},
  year = 2022,
  month = apr,
  journal = {Sensors},
  volume = {22},
  number = {8},
  pages = {2901},
  issn = {1424-8220},
  doi = {10.3390/s22082901},
  urldate = {2025-10-08},
  copyright = {https://creativecommons.org/licenses/by/4.0/},
  langid = {english}
}

@article{svedendahlRefractometricBiosensingBased2014,
  title = {Refractometric Biosensing Based on Optical Phase Flips in Sparse and Short-Range-Ordered Nanoplasmonic Layers},
  author = {Svedendahl, Mikael and Verre, Ruggero and K{\"a}ll, Mikael},
  year = 2014,
  month = nov,
  journal = {Light: Science \& Applications},
  volume = {3},
  number = {11},
  pages = {e220-e220},
  issn = {2047-7538},
  doi = {10.1038/lsa.2014.101},
  urldate = {2025-10-08},
  langid = {english}
}

@article{svedendahlRefractometricSensingUsing2009,
  title = {Refractometric {{Sensing Using Propagating}} versus {{Localized Surface Plasmons}}: {{A Direct Comparison}}},
  shorttitle = {Refractometric {{Sensing Using Propagating}} versus {{Localized Surface Plasmons}}},
  author = {Svedendahl, Mikael and Chen, Si and Dmitriev, Alexandre and K{\"a}ll, Mikael},
  year = 2009,
  month = dec,
  journal = {Nano Letters},
  volume = {9},
  number = {12},
  pages = {4428--4433},
  issn = {1530-6984, 1530-6992},
  doi = {10.1021/nl902721z},
  urldate = {2025-10-08},
  langid = {english}
}

@article{zitoLabelfreeDNABiosensing2021,
  title = {Label-Free {{DNA}} Biosensing by Topological Light Confinement},
  author = {Zito, Gianluigi and Sanit{\`a}, Gennaro and Guilcapi Alulema, Bryan and Lara Y{\'e}pez, Sof{\'i}a N. and Lanzio, Vittorino and Riminucci, Fabrizio and Cabrini, Stefano and Moccia, Maria and Avitabile, Concetta and Lamberti, Annalisa and Mocella, Vito and Rendina, Ivo and Romano, Silvia},
  year = 2021,
  month = nov,
  journal = {Nanophotonics},
  volume = {10},
  number = {17},
  pages = {4279--4287},
  issn = {2192-8614},
  doi = {10.1515/nanoph-2021-0396},
  urldate = {2025-10-08},
  copyright = {http://creativecommons.org/licenses/by/4.0},
  langid = {english}
}

@article{prabowoSurfacePlasmonResonance2018,
  title = {Surface {{Plasmon Resonance Optical Sensor}}: {{A Review}} on {{Light Source Technology}}},
  shorttitle = {Surface {{Plasmon Resonance Optical Sensor}}},
  author = {Prabowo, Briliant and Purwidyantri, Agnes and Liu, Kou-Chen},
  date = {2018-08-26},
  journal = {Biosensors},
  shortjournal = {Biosensors},
  volume = {8},
  number = {3},
  pages = {80},
  issn = {2079-6374},
  doi = {10.3390/bios8030080},
  url = {https://www.mdpi.com/2079-6374/8/3/80},
  urldate = {2025-11-11},
  langid = {english}
}

@article{capelliSurfacePlasmonResonance2023,
  title = {Surface Plasmon Resonance Technology: {{Recent}} Advances, Applications and Experimental Cases},
  shorttitle = {Surface Plasmon Resonance Technology},
  author = {Capelli, Davide and Scognamiglio, Viviana and Montanari, Roberta},
  date = {2023-06},
  journal = {Trends in Analytical Chemistry},
  shortjournal = {Trends in Analytical Chemistry},
  volume = {163},
  pages = {117079},
  issn = {01659936},
  doi = {10.1016/j.trac.2023.117079},
  url = {https://linkinghub.elsevier.com/retrieve/pii/S0165993623001668},
  urldate = {2025-11-11},
  langid = {english}
}

@article{luceTMMFastTransferMatrix2022,
  title = {{{TMM-Fast}}, a Transfer Matrix Computation Package for Multilayer Thin-Film Optimization: Tutorial},
  shorttitle = {{{TMM-Fast}}, a Transfer Matrix Computation Package for Multilayer Thin-Film Optimization},
  author = {Luce, Alexander and Mahdavi, Ali and Marquardt, Florian and Wankerl, Heribert},
  date = {2022-06-01},
  journal = {Journal of the Optical Society of America A},
  shortjournal = {J. Opt. Soc. Am. A},
  volume = {39},
  number = {6},
  pages = {1007},
  issn = {1084-7529, 1520-8532},
  doi = {10.1364/JOSAA.450928},
  url = {https://opg.optica.org/abstract.cfm?URI=josaa-39-6-1007},
  urldate = {2025-11-11},
  langid = {english}
}

@article{barolakLeveragingLowIndex2025,
  author       = {Barolak, Jonathan and Occhicone, Agostino and Finazzi, Marco and Biagioni, Paolo and Pellegrini, Giovanni},
  title        = {Leveraging Low Index Contrast to Reduce the Polarization Anisotropy in One-Dimensional Photonic Crystals},
  year         = {2025},
  eprint       = {2507.13193},
  archivePrefix= {arXiv},
  primaryClass = {physics.optics},
  note         = {arXiv:2507.13193},
}

@article{samocRefractiveindexAnisotropyOptical2007,
  title = {Refractive‐index Anisotropy and Optical Dispersion in Films of Deoxyribonucleic Acid},
  author = {Samoc, Anna and Miniewicz, Andrzej and Samoc, Marek and Grote, James G.},
  date = {2007-07-05},
  journal = {Journal of Applied Polymer Science},
  shortjournal = {J of Applied Polymer Sci},
  volume = {105},
  number = {1},
  pages = {236--245},
  issn = {0021-8995, 1097-4628},
  doi = {10.1002/app.26082},
  url = {https://onlinelibrary.wiley.com/doi/10.1002/app.26082},
  urldate = {2025-11-13},
  langid = {english}
}

@article{mishimaOpticalBirefringenceMultilamellar1996,
  title = {Optical Birefringence of Multilamellar Gel Phase of Cholesterol/Phosphatidylcholine Mixtures},
  author = {Mishima, Kiyoshi and Satoh, Koichi and Suzuki, Kiyomitsu},
  date = {1996-07},
  journal = {Colloids and Surfaces B: Biointerfaces},
  shortjournal = {Colloids and Surfaces B: Biointerfaces},
  volume = {7},
  number = {1--2},
  pages = {83--89},
  issn = {09277765},
  doi = {10.1016/0927-7765(96)01275-1},
  url = {https://linkinghub.elsevier.com/retrieve/pii/0927776596012751},
  urldate = {2025-11-13},
  langid = {english},
}

@article{keikhosraviRealtimePolarizationMicroscopy2021,
  title = {Real-Time Polarization Microscopy of Fibrillar Collagen in Histopathology},
  author = {Keikhosravi, Adib and Shribak, Michael and Conklin, Matthew W. and Liu, Yuming and Li, Bin and Loeffler, Agnes and Levenson, Richard M. and Eliceiri, Kevin W.},
  year = {2021},
  journal = {Scientific Reports},
  volume = {11},
  number = {1},
  pages = {19063},
  doi = {10.1038/s41598-021-98600-w},
  url = {https://www.nature.com/articles/s41598-021-98600-w}
}

@article{konopskyPhotonicCrystalSurface2007a,
  title = {Photonic {{Crystal Surface Waves}} for {{Optical Biosensors}}},
  author = {Konopsky, Valery N. and Alieva, Elena V.},
  date = {2007-06-01},
  journal = {Analytical Chemistry},
  shortjournal = {Anal. Chem.},
  volume = {79},
  number = {12},
  pages = {4729--4735},
  issn = {0003-2700, 1520-6882},
  doi = {10.1021/ac070275y},
  url = {https://pubs.acs.org/doi/10.1021/ac070275y},
  urldate = {2025-11-13},
  langid = {english}
}

@article{pellegriniChiralSurfaceWaves2017,
  title = {Chiral Surface Waves for Enhanced Circular Dichroism},
  author = {Pellegrini, Giovanni and Finazzi, Marco and Celebrano, Michele and Duò, Lamberto and Biagioni, Paolo},
  date = {2017-06-06},
  journal = {Physical Review B},
  shortjournal = {Phys. Rev. B},
  volume = {95},
  number = {24},
  pages = {241402},
  issn = {2469-9950, 2469-9969},
  doi = {10.1103/PhysRevB.95.241402},
  url = {http://link.aps.org/doi/10.1103/PhysRevB.95.241402},
  urldate = {2024-03-04},
  langid = {english}
}

@article{lechnerSituMonitoringRolling2021,
  title = {In situ {{Monitoring}} of {{Rolling Circle Amplification}} on a {{Solid Support}} by {{Surface Plasmon Resonance}} and {{Optical Waveguide Spectroscopy}}},
  author = {Lechner, Bernadette and Hageneder, Simone and Schmidt, Katharina and Kreuzer, Mark P. and Conzemius, Rick and Reimhult, Erik and Barišić, Ivan and Dostalek, Jakub},
  date = {2021-07-14},
  journal = {ACS Applied Materials \& Interfaces},
  shortjournal = {ACS Appl. Mater. Interfaces},
  volume = {13},
  number = {27},
  pages = {32352--32362},
  issn = {1944-8244, 1944-8252},
  doi = {10.1021/acsami.1c03715},
  url = {https://pubs.acs.org/doi/10.1021/acsami.1c03715},
  urldate = {2025-11-13},
  langid = {english}
}

@article{michelottiBlochSurfaceWaves2025,
  title = {Bloch Surface Waves on Dielectric One-Dimensional Photonic Crystals: Fundamental Properties and Applications [{{Invited}}]},
  shorttitle = {Bloch Surface Waves on Dielectric One-Dimensional Photonic Crystals},
  author = {Michelotti, Francesco},
  date = {2025-11-01},
  journal = {Optical Materials Express},
  shortjournal = {Opt. Mater. Express},
  volume = {15},
  number = {11},
  pages = {2839},
  issn = {2159-3930},
  doi = {10.1364/OME.568065},
  url = {https://opg.optica.org/abstract.cfm?URI=ome-15-11-2839},
  urldate = {2025-11-17},
  langid = {english}
}

@article{schmidtRollingCircleAmplification2022,
  title = {Rolling {{Circle Amplification Tailored}} for {{Plasmonic Biosensors}}: {{From Ensemble}} to {{Single-Molecule Detection}}},
  shorttitle = {Rolling {{Circle Amplification Tailored}} for {{Plasmonic Biosensors}}},
  author = {Schmidt, Katharina and Hageneder, Simone and Lechner, Bernadette and Zbiral, Barbara and Fossati, Stefan and Ahmadi, Yasaman and Minunni, Maria and {Toca-Herrera}, Jose Luis and Reimhult, Erik and Barisic, Ivan and Dostalek, Jakub},
  year = 2022,
  month = dec,
  journal = {ACS Applied Materials \& Interfaces},
  volume = {14},
  number = {49},
  pages = {55017--55027},
  issn = {1944-8244, 1944-8252},
  doi = {10.1021/acsami.2c14500},
  urldate = {2026-01-28},
  copyright = {https://creativecommons.org/licenses/by/4.0/},
  langid = {english}
}

@article{haslerDualElectronicOptical2025,
  title = {Dual {{Electronic}} and {{Optical Monitoring}} of {{Biointerfaces}} by a {{Grating-Structured Coplanar-Gated Field-Effect Transistor}}},
  author = {Hasler, Roger and Livio, Pietro A. and Bozdogan, Anil and Fossati, Stefan and Hageneder, Simone and {Montes-Garc{\'i}a}, Ver{\'o}nica and Movilli, Jacopo and Moazzenzade, Taghi and Loohuis, Luna and {Reiner-Rozman}, Ciril and Tamayo, Adri{\'a}n and Fiedler, Christine and Ib{\'a}{\~n}ez, Maria and Kleber, Christoph and Huskens, Jurriaan and Dostalek, Jakub and Samor{\`i}, Paolo and Knoll, Wolfgang},
  year = 2025,
  month = apr,
  journal = {IEEE Sensors Journal},
  volume = {25},
  number = {7},
  pages = {10521--10529},
  issn = {1530-437X, 1558-1748, 2379-9153},
  doi = {10.1109/JSEN.2025.3533113},
  urldate = {2026-01-28},
  copyright = {https://creativecommons.org/licenses/by/4.0/legalcode},
  langid = {english}
}

@article{anopchenkoEffectThicknessDisorder2016a,
  title = {Effect of Thickness Disorder on the Performance of Photonic Crystal Surface Wave Sensors},
  author = {Anopchenko, Aleksei and Occhicone, Agostino and Rizzo, Riccardo and Sinibaldi, Alberto and Figliozzi, Giovanni and Danz, Norbert and Munzert, Peter and Michelotti, Francesco},
  year = 2016,
  month = apr,
  journal = {Optics Express},
  volume = {24},
  number = {7},
  pages = {7728},
  issn = {1094-4087},
  doi = {10.1364/OE.24.007728},
  urldate = {2026-01-28},
  copyright = {https://doi.org/10.1364/OA\_License\_v1\#VOR-OA},
  langid = {english}
}

@article{nallurSignalAmplificationRolling2001,
  title = {Signal Amplification by Rolling Circle Amplification on {{DNA}} Microarrays},
  author = {Nallur, G.},
  year = 2001,
  month = dec,
  journal = {Nucleic Acids Research},
  volume = {29},
  number = {23},
  pages = {118e-118},
  issn = {13624962},
  doi = {10.1093/nar/29.23.e118},
  urldate = {2026-01-29},
  langid = {english}
}

@article{xiangRealtimeMonitoringMycobacterium2015,
  title = {Real-Time Monitoring of Mycobacterium Genomic {{DNA}} with Target-Primed Rolling Circle Amplification by a {{Au}} Nanoparticle-Embedded {{SPR}} Biosensor},
  author = {Xiang, Yang and Zhu, Xiaoyan and Huang, Qing and Zheng, Junsong and Fu, Weiling},
  year = 2015,
  month = apr,
  journal = {Biosensors and Bioelectronics},
  volume = {66},
  pages = {512--519},
  issn = {09565663},
  doi = {10.1016/j.bios.2014.11.021},
  urldate = {2026-02-03},
  langid = {english}
}

@article{huangProteinDetectionTechnique2007,
  title = {A Protein Detection Technique by Using Surface Plasmon Resonance ({{SPR}}) with Rolling Circle Amplification ({{RCA}}) and Nanogold-Modified Tags},
  author = {Huang, Yi-You and Hsu, Hsin-Yun and Huang, Chi-Jer Charles},
  year = 2007,
  month = jan,
  journal = {Biosensors and Bioelectronics},
  volume = {22},
  number = {6},
  pages = {980--985},
  issn = {09565663},
  doi = {10.1016/j.bios.2006.04.017},
  urldate = {2026-02-03},
  langid = {english}
}







\end{document}